\documentclass[pra,aps,twocolumn]{revtex4-1}
\usepackage[pdftex]{graphicx}

\begin{document}
\begin{flushleft}
\textbf{Comment on ``Two Fundamental Experimental Tests of Nonclassicality with Qutrits``}
\end{flushleft}

In a recent paper \protect\cite{Ahrens} Ahrens \textit{et al.} claim that our ``(...)\ experiment on qutrits does not test Klyachko \textit{et al.}'s inequality \protect\cite{Klyachko}, but an inequality with extra correlations'' and that the ``(...)\ experiment cannot be considered a proper test of a noncontextuality inequality, since the same observable is measured with different setups in different contexts''. We disagree with these claims. In this note we briefly re-state how our test of the non-contextuality inequality was constructed \protect\cite{Lapkiewicz}. We explain how we keep the context of measurements when switching between different terms of the tested inequality, and we argue that we did in fact test the Klyachko, Can, Binicioglu and Shumovsky (KCBS) inequality.  In doing so, we also clarify why our experiment is indeed a proper test of noncontextual realism.

Ahrens \textit{et al.}\ disregard the way our setup is constructed and this leads them to their incorrect conclusions about our experimental test of the KCBS inequality \protect\cite{Lapkiewicz}. We stress that we do not base our argument on the mathematical description of the setup, but instead, we physically enforce the fulfillment of the requirements of the test of a non-contextuality inequality: we always keep one of the measurements physically fixed while changing its accompanying measurement. Thus we explicitly test whether the measurement which is kept fixed depends on the context of the accompanying measurement.  Below we explain our experiment in more detail.

The starting point of our setup is a single-photon source which provides us with a heralded photon in a well-defined spatial mode. At the other end of the setup we have detectors monitoring two out of three distinct modes. To test the KCBS inequality \cite{Klyachko}:
\begin{equation}
\label{kcbs}
	\left\langle A_1 A_2 \right\rangle + \left\langle A_2 A_3 \right\rangle + \left\langle A_3 A_4 \right\rangle + \left\langle A_4 A_5 \right\rangle + \left\langle A_5 A_1 \right\rangle  \geq -3,
\end{equation}
we need to be able to perform two-outcome measurements. We define the measurement outcomes in the following way: If a detector clicks we assign $-1$ to the corresponding observable, if it does not click we assign $+1$ to it. 

\begin{figure}[!t]
 
	\centering
		\includegraphics[width=.45\textwidth]{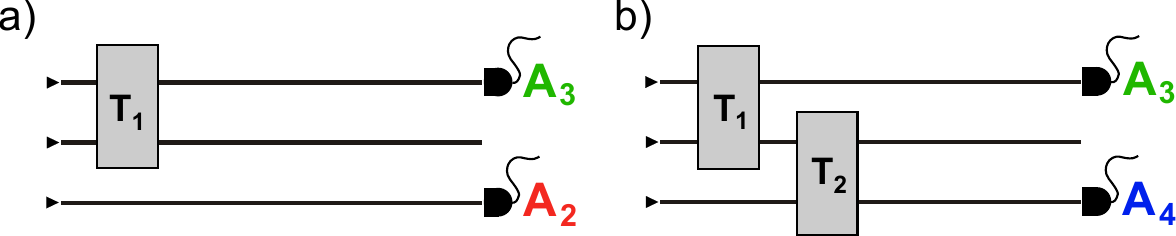}
	\caption {Two (out of five) measurement stages. Straight, black lines represent the optical modes (beams), gray boxes represent transformations on the optical modes. A key aspect of our experimental implementation is that each transformation acts only on two modes, leaving the other mode completely unaffected. Thus, the part of the physical setup corresponding to the measurement of $A_3$ is exactly the same in both (a) and (b) (likewise for other measurements which are not shown here).}
	\label{fig:boxes}
\end{figure}

Fig.\ref{fig:boxes} shows two out of five different configurations of the measurement setup used to measure the five terms of the KCBS inequality. When moving from the setup depicted in Fig.\ref{fig:boxes}.a to that of Fig.\ref{fig:boxes}.b, we apply a transformation to the lower two modes and leave the uppermost mode untouched.  The crucial property of our experiment is that we physically keep one of the measurements unchanged when moving between the terms of the inequality. In other words, the measurement common to both configurations is not just alike in the two cases, but is in fact the identical physical measurement. Mathematically, this can be described by a rotation whose axis is the vector assigned to the unaffected mode. To move between the further terms of the inequality, we keep rotating the basis in the same way, always keeping one of the measurements (vectors we project onto) unchanged. 
Nevertheless, the last measurement is not physically the same as $A_1$, from which we started. They would be identical in the ideal case, in which the transformations would be performed perfectly, but even in this case, these would be two different physical realisations of the same projector. 
To account for this, we replace $A_1$ in the last term of the KCBS inequality with $A_1'$ and add one extra term to obtain: 
\begin{equation}
\label{eq:Km}
	\left\langle A_1 A_2 \right\rangle + \left\langle A_2 A_3 \right\rangle +  \left\langle A_3 A_4 \right\rangle + \left\langle A_4 A_5 \right\rangle + \left\langle A_5 A_1' \right\rangle - \overline{A_1' A_1} \geq -4.
\end{equation}
The derivation of the modified inequality is analogous to that of the original KCBS inequality. 
We also note here that it is the topology of the original KCBS diagram, not of our particular implementation, that is the reason why the extra correlation needs to be measured. This measurement certifies that after traversing the KCBS diagram we indeed come back sufficiently close to the starting point. 

To measure $\overline{A_1' A_1}$ we need to find experimentally when the two measurements have the same result, and when not. Physically this can be viewed as checking how well the corresponding modes overlap. If the modes were the same, the two measurements would be identical and have always the same results. (In this ideal case $\overline{A_1' A_1}=1$, and the inequality (\ref{eq:Km}) reduces to the original KCBS inequality.) This would mean that whenever we block one mode in $A_1$, we perfectly block the same mode in $A_1'$ also. We base our method of measurement of $\overline{A_1' A_1}$ on this observation. For more details see \protect\cite{Lapkiewicz}.

It is crucial for experimental tests of the KCBS inequality that there is one measurement common to neighbouring pairs of measurements (e.g. the same physical realisation of the measurement $A_3$ appears in the measurements of pairs $\left\langle A_2 A_3 \right\rangle$ and $\left\langle A_3 A_4 \right\rangle$). As we discussed above, this requirement is met by our experiment intrinsically, because we leave one of the modes (and therefore the corresponding projector) completely unchanged when moving between the terms of the KCBS inequality.

This work was supported  by  the  European  Research  Council  (advanced  grant  QIT4QAD,  227844), the
Austrian  Science  Fund  (FWF)  through  the  Special  Research  Program  (SFB)  Foundations  and
Applications  of  Quantum  Science  (FoQuS, F4006-N16), and by the European Commission (SIQS, FP7 600645). M.W. acknowledges project quasar from the National Center of Research and
Development (NCBiR) of Poland and project HOMING PLUS from the Foundation
for Polish Science.

\small R. Lapkiewicz$^{1,2}$, P. Li$^{3}$, C. Schaeff$^{1,2}$, N. K. Langford$^{4}$, S. Ramelow$^{1,2}$, M. Wie\'sniak$^{5}$, A. Zeilinger$^{1,2,6}$. \newline

$^{1}$ Quantum Optics, Quantum Nanophysics, Quantum Information, University of Vienna, 
Boltzmanngasse 5, Vienna A-1090, Austria 
$^{2}$ Institute for Quantum Optics and Quantum Information, Austrian Academy of Science, 
Boltzmanngasse 3, Vienna A-1090, Austria 
$^{3}$ Columbia University, New York, New York 10027, USA
$^{4}$ Department of Physics, Royal Holloway, University of London, London TW20 0EX, UK 
$^{5}$ Institute of Theoretical Physics and Astrophysics, University of Gda´nsk, 80-952 Gda´nsk, Poland 
$^{6}$ Vienna Center for Quantum Science and Technology, Faculty of Physics, University of Vienna, Boltzmanngasse 5, Vienna A-1090, Austria 


\begin{thebibliography}{99}
\bibitem{Ahrens} J. Ahrens, E. Amselem, A. Cabello, M. Bourennane, quant-ph/1301.2887 (2013).
\bibitem{Klyachko} A.A. Klyachko, M. Ali Can, S. Binicioglu, and A. S. Shumovsky, {\em Phys. Rev. Lett.} {\bf 101}, 020403 (2008).
\bibitem{Lapkiewicz} R. Lapkiewicz, P. Li, C. Schaeff, N. K. Langford, S. Ramelow, M. Wie\'sniak, and A. Zeilinger, {\em Nature} {\bf 460}, 490 (2011).
\end{thebibliography}
\end{document}